\documentclass[12pt]{article}

\usepackage{graphicx}
\usepackage{wrapfig}
\usepackage{times}
\usepackage{natbib}
\usepackage{amsmath,amssymb,amsfonts}

\makeatletter
\def\la{\mathrel{\mathpalette\gl@align<}}
\def\ga{\mathrel{\mathpalette\gl@align>}}
\def\gl@align#1#2{\lower.6ex\vbox{\baselineskip\z@skip\lineskip\z@\ialign{$\m@th#1\hfil##\hfil$\crcr#2\crcr\sim\crcr}}}
\makeatother

\title{An Overview of the Square Kilometre Array}
\author{\parbox{\textwidth}{Minh T Huynh$^{1, 2}$, Joseph Lazio$^{3,2}$\\
{\small 1.\ International Center for Radio Astronomy Research, University of Western Australia, Stirling Highway, Crawley, Western Australia 6009, Australia} \\
{\small 2.\ SKA Organisation,  Jodrell Bank Observatory, Cheshire SK11 9DL, United Kingdom }\\
{\small 3.\ Jet Propulsion Laboratory, California Institute of Technology, 4800 Oak Grove Drive, Pasadena, CA 91106, USA}
}}

\date{} %Leave this blank

\begin{document}

\maketitle

\begin{abstract}

The Square Kilometre Array (SKA) will be the premier instrument to study radiation at centimetre and metre wavelengths from the cosmos, and in particular hydrogen, the most abundant element in the universe. The SKA will probe the dawn of galaxy formation as well as allow advances in many other areas of astronomy, such as fundamental physics, astrobiology and cosmology. Phase 1, which will be about 10\% of the full SKA collecting area, will be built in Australia and South Africa. 
This paper describes the key science drivers of the SKA, provides an update on recent SKA Organisation activities and summarises the baseline design for Phase 1.

\end{abstract}

\section{Introduction}

In the 21st Century, we seek to understand the Universe that we inhabit. In the latter half of the 20th
Century, we discovered both unimagined sources and phenomena. Observations at radio
wavelengths laid the foundation for many of these discoveries, including non-thermal emission
mechanism, active galaxies, the cosmic microwave background (CMB), pulsars, gravitational
lensing, and extrasolar planets.

Currently under development by an international consortium, the Square Kilometre Array (SKA) is
a centimeter- and meter-wavelength telescope envisioned as being one of a suite of multiwavelength
facilities for the 21st Century. The original motivation for the SKA was a Òhydrogen
array,Ó a telescope with enough sensitivity to detect the 21-cm H~I line from a Milky Way-like
galaxy at a redshift of order unity. Since that time, it has been recognized that the science program
that could be carried out with such a telescope is much broader, and the international community
has developed a set of Key Science Programs of the SKA that are intended to address a broad range
of fundamental questions in astronomy, physics, and astrobiology.

In this paper, we describe the Key Science Programs of the SKA (Carilli and Rawlings 2004;
Gaensler et al.~2003). In the spirit of the European Astronet Roadmap process and the U.S. New Worlds,
New Horizons Decadal Survey, we shall divide the key science into two broad categories, ÒoriginsÓ
and Òfundamental physics,Ó recognizing that the divisions between these two categories may, at
times, be indistinct.

%Example how to use figre environment
%\begin{figure}[tbhp]
%\centering\includegraphics[width=0.7\textwidth]{}
%\caption{}\label{}
%\end{figure}

%Example how to use wrapfigure environment
%\begin{wrapfigure}{r}{65mm}
%\vspace{-1cm}
%\centering\includegraphics[width=6cm]{}
%\caption{}\label{}
%\end{wrapfigure}

\section{Key Science: Origins}

One of the motivations for observing the Universe is that it can answer fundamental questions about
how we originated, questions that have been posed since (or before!) the beginning of humanity.

\subsection{Emerging from the Dark Ages and the Epoch of Reionization}

The focus of this Key Science Program is the first luminous objects in the Universe and their
formation. At a redshift around 1100, the Universe became largely neutral as protons and electrons
combined to form the first hydrogen atoms and the photons that we now see as the cosmic
microwave background (CMB) began free streaming across the Universe. Today the Universe is largely
ionized. The epoch of reionization of the Universe is dated to z $\sim$ 6 to 11 from 
observations of the Gunn-Peterson trough in spectra of high redshift quasars (Fan et al.~2000) and the analysis
of observations of the CMB from Wilkinson Microwave Anisotropy Probe (WMAP) (Spergel et al.~2003;
Komatsu et al.~2011) and Planck (Planck Collaboration 2013).
The interval from a redshift of about 1100 to the epoch of reionization is
known as the Dark Ages and Cosmic Dawn. The redshift of the epoch of reionization is so large that
only observations at wavelengths longer than 1 micron are useful, and the SKA will play a key role
in probing two aspects of the end of the Dark Ages.

First, as the first structures form and the first stars and accreting black holes begin to illuminate
their surroundings, they should ionize and heat the surrounding hydrogen gas in the intergalactic
medium (IGM). Its excitation (spin temperature) will decouple from the temperature of the cosmic
microwave background, and a complex, time-dependent patchwork of (highly redshifted) hydrogen
emission or absorption against the background CMB on the sky is predicted to result (see Pritchard
and Loeb 2012, and references therein). The simplest result is that Stromgren spheres will be
formed around the first stars, the first accreting black holes (quasars), or both. More complicated
scenarios involve the heating of the hydrogen gas as it collapses even before the first stars form.
The current constraints on the redshift at the epoch of reionization indicate that the 21-cm (1420
MHz) emission from the remaining neutral hydrogen will be visible at wavelengths between about
1.5 and 2.5 meters (frequencies between about 120 and 200 MHz). The goal of the SKA is to detect
this highly-redshifted neutral hydrogen emission or absorption, which in turn will constrain the
formation of the first structures.

Any radio-loud objects that exist before Reionization is complete should display 21-cm absorption
features in their radio spectra (Furlanetto and Loeb 2002; Mack and Wyithe 2012). Analogous to the Lyman-alpha forest, the so-called 21-cm forest
would be a powerful probe of small-scale structures before Reionization, albeit only on lines of
sight to radio-loud objects. There is a degeneracy in 21cm tomography between cosmological and
astrophysical parameters when H~I spin temperature rises above the CMB temperature (Santos and
Cooray 2006). The 21 cm forest observations are complementary to 21cm tomography because they
help us determine the spin temperature at the relevant redshifts, reducing this degeneracy.

The Experiment to Detect the Global EOR Signature (EDGES) is a single antenna experiment at the Western Australian SKA site 
designed to constrain the global signature of the EoR. The EDGES all-sky spectrum between 100 and 200 MHz 
is able to exclude rapid reionization on timescales of $\delta z$ $<$ 0.06 (Bowman and Rogers 2010). The power spectrum from the Precision Array for Probing the Epoch of Reionization (PAPER) 32 tile deployment in South Africa has a 2$\sigma$ limit of (52mK)$^2$ for $k = 0.11$ $h$ Mpc$^{-1}$ at $z$ = 7.7 (Parsons et al.~2013), which implies there is X-Ray heating of the IGM (for certain reionization models).  
The recently completed Murchison Widefield Array (MWA) has begun observations with 128 tiles and a detection (SNR $>$ 1) of the EoR power spectrum (for  $k < 0.08$ $h$ Mpc$^{-1}$) is possible in 1000 hours of observations (Thyagarajan et al. 2013). These are examples of experiments on the Square Kilometre Array sites which are already providing data on the epoch of reionization. 
The goal of the SKA is to fully image the IGM across a wide range of redshift to observe the the universe's transition from a neutral to ionized state. 
 
 \subsection{Galaxy Formation and Evolution}

The original focus of the SKA was observations of the 21-cm H~I line from galaxies, and such
observations remain a significant focus of the SKA Key Science Program. Neutral hydrogen is the
raw material from which stars form. The peak of the star formation rate in the Universe occurred at
redshifts between about 1 to 2 (e.g. Madau et al.~1996; Steidel et al.~1999; Hopkins and Beacom
2006). The SKA will be able to probe the evolution of neutral hydrogen to this crucial point in the
assembly of galaxies.

Although the star formation rate in the Universe peaks at redshifts of 1 -- 2, the density of H~I in the
Universe appears to be quite constant until relatively recently (Lah et al.~2007). 
Limits on the cosmic H~I density have large uncertainties and largely rely on stacking, 
but they indicate the drop off in gas content occurred only in the last few Gyrs (Delhaize et al.~2013).
Moreover, the amount of gas that most galaxies contain is insufficient to power their star formation for a Hubble
time (Schiminovich et al.~2010). These observations suggest that galaxies are able to tap a
reservoir of gas in order to power their star formation. This reservoir might be the IGM, with hot
gas condensing onto galaxies or delivered through ``cold accretion" streams or it might be in the
form of mergers or both. The very deepest observations often show extended H~I halos around
galaxies (e.g. de Blok et al.~2008) or low column density clouds in the neighbourhoods of galaxies
(Shull et al.~1996; Putman et al.~2003). It is also well known that the H~I gas is often an indicator of
potential merger activity in groups of galaxies. However, the current generation of observations is
not sufficiently deep to provide definitive answers to the question of the balance between accretion
and mergers in the galactic gas budget.

Simulations of the formation of late-type spirals, such as the ERIS simulation (Guedes et al.~2011),
are beginning to successfully reproduce galaxy structural properties, total H~I mass, H~I-to-stellar
mass ratios and cold-vs-hot phase ratios which are all consistent with a host of observational
constraints. While further work needs to be done on these simulations, including exploring different
merger histories, gas acquired through ``cold mode accretion" appears to be essential in the
formation of galaxies (e.g. Keres et al.~2009). Observations by the SKA will inform and confirm the
physics underlying these simulations.

\subsection{Astrobiology: The cradle of life}

The current heuristic picture of planetary assembly, supported by considerable evidence from our
solar system and young stars in star formation regions, is that it begins in a disk composed of dust
and gas. The initial dust grain size is probably sub-micron, comparable to that for interstellar dust
particles. Within the proto-planetary disk, the dust grains begin to ``stick" together. As they do so,
they decouple from the disk gas and begin to interact gravitationally. The dust grains continue to
accrete, forming ``pebbles", then ``boulders", and finally planetesimals (e.g. Lissauer 1993). One
difficulty with this scenario is, given their kinetic energies, how dust grains accrete to form
``pebbles" rather than destroying each other. Probing this crucial regime of planet
formation requires observations at wavelengths comparable to the size of the particles, so on scales
of order 1 cm. With its high frequency capabilities (observations to 25 GHz or 1.2 cm wavelength),
the SKA will be positioned uniquely to probe the assembly of planets. Moreover, it is planned for
the SKA to be able to obtain milliarcsecond resolution. At the distance of nearby star forming
regions ($\sim$ 150 pc, e.g., Taurus, Ophiucus, and Chamaeleon), 1 AU subtends an angle of
approximately 7 milliarcseconds. Thus, the SKA will be able to resolve the inner portions of protoplanetary
disks. For a solar-mass star, the orbital period at a distance of 1 AU is 1 year, so the SKA
will even be able to make synoptic observations of the inner regions of proto-planetary disks,
potentially assembling ``movies" of planet formation.

In addition, a number of large ($>$ 10 atom) prebiotic molecules are being discovered in interstellar
space (Hollis et al.~2006; Remijan et al.~2008; Zaleski et al.~2013). Typical transition frequencies for these molecules lie
in the 10 to 20 GHz range, and the expectation is that even larger molecules would have transitions
at lower frequencies. The SKA will search for these prebiotic molecules and explore the extent of
organic chemistry and the precursors of life in interstellar space. Finally, detecting transmissions
from another civilization would provide immediate and direct evidence of life elsewhere in the
Universe. The ``Waterhole" is the frequency regime bounded by H~I (1420 MHz) and the OH hydroxyl ion (1665, 1667 MHz), 
and some have argued that if another civilization exists they would transmit between these well known natural emission lines. 
With its sensitivity, not only will the SKA probe deeper into the Galaxy than any previous survey, for the first time it will enable searches at power levels comparable to that of terrestrial TV transmitters. 

\section{Key Science: Fundamental Physics}

A second motivation for astronomy is that it can motivate or provide tests of theories of
fundamental physics. For instance, observations of gravitational lensing by the Sun provided some
of the key early support for the EinsteinÕs General Theory of Relativity (GR), which now finds
widespread use, such as in corrections applied to timing signals from the Global Positioning System
(GPS) satellites.

\subsection{Strong Field Tests of Gravity using Pulsars and Black Holes}

Observations of a particular pulsar, PSR B1913+16, have already provided an indirect detection of
the gravitational radiation predicted in GR, as well as the 1993 Nobel Prize in Physics (Taylor and
Weisberg 1989). However, the Galaxy should contain a number of systems capable of providing
even more stringent tests. The focus of this Key Science Project is to conduct a census of the
Galaxy for radio pulsars and identify those objects best suited for probing strong field gravity within
the context of theories of gravity such as GR. Current estimates are that the Milky Way Galaxy
contains about 20,000 rotation-powered radio pulsars beamed in our direction (Smits et al.~2009).
The sensitivity of the SKA will be such that it should be able to detect a significant fraction of these
pulsars. After completing the census of pulsars, observations with the SKA will focus on high
precision timing observations of these pulsars, with an aim toward two aspects of GR.

First, population studies of neutron stars and radio pulsars suggest that there should be at least one
pulsar-black hole binary system and potentially as many as 100 in the Galaxy. The pulses from a
pulsar can be considered to form a clock. As a black hole is the most compact object that should
exist, a clock in its environment would provide stringent tests of various predictions of GR. At a
basic level, the pulsar timing will reveal the properties of the black hole companion, such as its
mass and angular momentum, in a manner similar to how pulsar timing observations have measured
the mass of both components in double neutron star systems. High precision timing will allow
higher order tests to be conducted, such as tests of the ``no-hair" theorem that predicts that a black
hole is described entirely by its mass, angular momentum, and electric charge, and which also
predicts a simple relation between its angular momentum and quadrupole moment. Other tests that
could be conducted include searching for possible violations of the strong equivalence principle or
evidence for gravitational theories beyond GR (e.g., tensor-scale theories). An important aspect of
this search is the Galactic center. The supermassive black hole in the Galactic center should contain
a number of pulsars in orbit about it, pulsars which have escaped detection because current
instruments do not have the necessary sensitivity at the frequencies ($>$ 10 GHz) required to mitigate
the severe interstellar scattering effects along the line of sight.

Second, the SKA is expected to discover a network of millisecond pulsars across the sky. With
their exquisite timing stability, millisecond pulsars are amongst the most accurate clocks available.
Effectively, this network of millisecond pulsars (the pulsar timing array) can serve as a many-armed
gravitational wave detector, searching for timing distortions due to the passage of low-frequency
gravitational waves ($\sim$ nHz). Generally, cosmic sources are expected to produce a spectrum of
gravitational waves, and the SKA pulsar timing array will probe gravitational waves produced
either by single sources such as supermassive binary black holes (Jenet et al.~2004) or a stochastic
background of binary supermassive black holes (Jaffe and Backer 2003), cosmic strings or relic GWs
from the big bang (Maggiore 2000).

There are currently three major pulsar timing consortia, the Parkes Pulsar Timing Array (\hbox{PPTA}, Yardley et al.~2010; Shannon et al.~2013), the European Pulsar Timing Array (\hbox{EPTA}, van~Haasteren et al.~2011), and the North American Nanohertz Observatory for Gravitational Waves (NANOGrav, Demorest et al.~2013).  Together these programmes are obtaining high precision measurements of the times of arrival from about 40 pulsars.  To date, none of these programmes have detected the stochastic background from the ensemble of supermassive black hole binaries, but their sensitivities are beginning to encroach substantially on the expected levels.  
This emerging tension between observations and theoretical expectations demands continuing observations, such as the SKA will provide, and may ultimately require substantial modifications to our understanding of the end stages of galaxy mergers or the generation of gravitational waves.

\subsection{The Origin and Evolution of Cosmic Magnetism}

The focus of this Key Science Project is an all-sky grid of magnetic field measurements from which
to probe the role and origin of the magnetic field in the Galaxy, galaxy clusters, and intergalactic
space. Electromagnetism is one of the most accurate physical theories, and it is clear that magnetic
fields fill intracluster (e.g. Xu et al.~2006) and interstellar space (e.g. Simard-Normandin and
Kronberg 1979). Magnetic fields affect the evolution of galaxies and galaxy clusters, contribute
significantly to the total pressure of interstellar gas, are essential for the onset of star formation, and
control the density and distribution of cosmic rays in the interstellar medium. Nonetheless, fairly
basic questions remain about the origin and evolution of cosmic magnetic fields. A radio wave
propagating through a magnetized plasma undergoes Faraday rotation, providing the SKA with a
unique probe of cosmic magnetic fields.

The tool by which the SKA will probe cosmic magnetic fields is an all-sky Faraday rotation
measure survey. The largest current survey of magnetic field measurements comprises rotation measures of 4000 extragalactic radio sources 
(Hammond et al.~2012), but the SKA is expected to be able to measure the Faraday rotations of order $2 \times 10^7$
extragalactic sources (and perhaps all 20,000 pulsars in the Galaxy). This all-sky grid will provide
a typical separation of about 90 arcseconds between magnetic field measurements.

With this grid, a detailed picture of the Galactic magnetic field will be produced, and similar
measurements will be used to probe the fields in nearby galaxies. Such a detailed model for
galactic magnetic fields can in turn discriminate between various origins for magnetic fields in
galaxies, whether the fields are in some sense primordial or were generated at later times by a
dynamo action (e.g., a so-called $\alpha$-$\Omega$ dynamo, Beck et al.~1996).

For relatively nearby clusters of galaxies, the grid of magnetic field measurements will be
sufficiently dense to probe the field within the clusters themselves, in contrast to the current
situation in which only properties averaged over many clusters can be determined. A detailed view
of the magnetic field structure within clusters will in turn allow probes of the interaction between
magnetic fields and the hot X-ray emitting gas, as well as the interplay between various ``heating"
mechanisms for a cluster (e.g., mergers, radio jets from active galactic nuclei near the center of
clusters) and the cooling provided by the X-ray emission.

Finally, with the deepest SKA observations, magnetic field measurements at high redshift ($z > 2$)
will be possible. Complementing the field measurements in nearby galaxies, observations of the
field in distant galaxies may trace directly the enhancement of the field by a dynamo (or illustrate
that a dynamo is not responsible for the origin of the field). Also, the SKA all-sky grid will allow
searches for any intergalactic magnetic field (i.e., one outside of clusters and permeating
intergalactic space as a whole). Such an intergalactic magnetic field may have played a role in the
assembly of large-scale structure and the formation of the ``cosmic web".

\subsection{Cosmology and Dark Energy}

The H~I emission from galaxies can be used to study the galaxies themselves, or it can be used to
identify test masses from which one can conduct cosmological observations. If the SKA can
observe the H~I emission from galaxies out to a redshift of order unity over much of the sky ($\sim$ 2$\pi$
sr), it will survey a significant volume of the Universe ($\sim$ 100 Gpc$^3$). Within this volume should be
more than one billion galaxies, and the galaxy power spectrum as a function of redshift can be
determined. At the time of recombination, acoustic oscillations in the intergalactic plasma should
have been ``frozen in" oscillations that today would reveal themselves as baryon acoustic
oscillations (BAOs) in the galaxy power spectrum (Eisenstein and Hu 1998). These BAOs can be
considered as a standard ruler (e.g. Blake and Glazebrook 2003), and the SKAÕs sensitivity should be
such that they can be determined as a function of redshift. The SKA experiment will determine the
change in the apparent angular size of these acoustic oscillations as a function of redshift. When
combined with measurements of the size of these oscillations seen in the CMB, one can obtain a
measure of the cosmic evolution of the Universe. In particular, the influence of dark energy from
the time of the formation of the CMB to $z \sim 1$ can be probed, thereby constraining the equation of
state of the Universe. Crucially, the accuracy of measurements of this sort depends upon the total
number of objects detected. The large sample size of the SKA surveys will provide unparalleled
precision.

An alternate approach to dark energy studies with H~I is intensity mapping. The relevant scale for
BAOs is about 150 Mpc, much larger than the size of an individual galaxy, or even a group of
galaxies. Thus, rather than attempting to resolve individual galaxies, intensity mapping seeks to
detect the integrated emission from galaxies. Once the integrated emission is detected, however, the
approach to BAO studies of dark energy is conceptually similar in that the objective is to detect
BAOs via a power spectrum analysis of the H~I emission.

The SKA will also be able to explore weak lensing.  Strong lensing is where a massive body can distort light rays from a single distant source sufficiently so that multiple images are seen by observers, and extreme examples of this are Einstein rings (e.g. MG 1131+0456, Hewitt et al.~1988). In weak lensing the bending of light is so small that the shape of the galaxy is distorted by only a small amount, on the order of 0.1\% for weak lensing by large-scale structure (cosmic shear). Therefore weak lensing is hard to detect for an individual galaxy but it can be detected statistically with a large enough ensemble of galaxies using, for example, two and three-point shear correlation functions (e.g. Bartelmann and Schneider 2001). Weak lensing measurements will provide powerful constraints on dark energy and modified gravity, the mass power spectrum, and information on the distribution and mass profiles of galaxy and cluster halos. Its importance is that it can provide constraints in some sense ``orthogonal" to those provided by BAO measurements. A modest angular resolution ($\lesssim 0.5$ arcsec) all-sky survey with the SKA would be a powerful measure of weak lensing. To fully realise the potential of weak lensing requires controlling systematics, and the major systematic which affects the measurement of galaxy shapes is the telescope point spread function (PSF). Optical experiments have to contend with many contributions to the PSF, including atmospheric seeing, telescope optics, and non-flatness of the focal and CCD plane, whereas to first order, the PSF of a radio telescope array should be determined merely by the configuration of the individual telescopes, which is known to extremely high accuracy.

\vspace{-2mm}
\section{Discovery Space}
\vspace{-1mm}

The history of science has shown that new technologies lead to discoveries. The field of radio
astronomy itself is an excellent example of this phenomenon. Celestial radio emission was
discovered serendipitously by Karl Jansky during his investigations into the sources of static in
long-distance radio transmissions. As mentioned above, since its discovery, observations at radio
wavelengths have laid much of the groundwork for modern astronomy, including the discovery of
non-thermal emission processes, quasars, the cosmic microwave background, pulsars, masers, and
extrasolar planetary systems. Moreover, many of these discoveries were themselves serendipitous.
The time domain has proven to be a rich source of discoveries, from pulsars (Hewish et al.~1968) to the more recent fast radio bursts (FRBs, Thornton et al.~2013).  The latter appear to be a previously unrecognized population of extragalactic radio emitters and illustrate the potential for future discoveries.

\section{Project Updates}

\vspace{-2mm}
\subsection{Phased Approach}
\vspace{-1mm}

One of the key strengths of an interferometer is that, if receptors are removed from operation, the
interferometer degrades gracefully. Conversely, an interferometer can begin science operations
well before it has reached its full complement of receptors. Such is the notion of the SKA Phase 1:
rather than waiting for the construction of the full SKA to be completed, significant science results
can already be produced when the array has only a fraction of its full complement of receptors. 
A phased approach can also reduce risk by adopting technically mature and well-costed 
engineering solutions. A notional value for the scale of SKA Phase 1 is that it would be 10\% of the capability of SKA (Garrett et al.~2010). Key
scientific motivations for SKA Phase 1 are two aspects of the full SKA Science Case, namely,
studies of H~I over cosmic time, particularly observations of the H~I in the IGM during the Epoch of
Reionization, and fundamental physics as probed by pulsar observations, particularly detecting
gravitational waves.

\vspace{-2mm}
\subsection{The SKA Organisation}
\vspace{-1mm}

The office of the SKA Organisation, located at Jodrell bank Observatory near Manchester, England, is responsible for co-ordinating the global work of the SKA project. 
In  2011 the SKA Organisation was formally established as a private company in the United Kingdom, limited by guarantee. 
Seven countries were initial members: Australia, China, Italy, the Netherlands, New Zealand, South Africa, and the UK. 
Every full SKA member appoints two representatives to the Board of Directors, one voting and one non-voting.
The Board of Directors oversees the work of the SKA Organisation. It has the authority to appoint senior staff, decide budgets, 
admit new members to the organisation and direct the work of the global work package consortia. 
Canada joined in March 2012, Sweden in June 2012, and Germany in December 2012. India is an associate member and is expected to become a full member shortly. 

\vspace{-2mm}
\subsection{Site Selection}
\vspace{-1mm}

To maximise the scientific returns of the SKA it is essential that it is built at the best possible site.
The general required physical characteristics of the host site are that it must be very radio quiet, be
at least $\sim$3000 km in extent to allow for long baselines, and have very low ionospheric and
tropospheric turbulence. In 2006, two candidate sites were short-listed to host the SKA: South
Africa and Australia. The Australia and New Zealand proposal is to have the core at the Murchison
Radio-astronomy Observatory, approximately 300 km north east of Geraldton in Western Australia.
The peripheral stations in New Zealand would then allow for baselines as great as 5000 km. The
South African site is located in the Karoo region of the Northern Cape province of South Africa,
with stations located across eight countries on the continent: Namibia, Botswana, Mozambique,
Zambia, Mauritius, Madagascar, Kenya and Ghana. Both sites have had significant infrastructure
development and are hosts to the upcoming SKA precursors ASKAP (Australia) and MeerKAT
(South Africa).

In May 2012 the members of the SKA Organisation announced the outcome of the site selection process. They decided on a dual-site implementation to maximise the value of the investments made on both sites and to maintain an inclusive approach to the SKA. The agreement means that in Phase 1 Australia will host the low frequency aperture arrays (SKA1-Low, see Figure 1) and a survey instrument that incorporates the 36 ASKAP dishes and approximately 60 additional dishes with phased array feeds (SKA1-Survey). In Phase 1 South Africa the SKA will incorporate the 64 dish MeerKAT with about another 190 dishes equipped with sensitive single pixel feeds into a sensitive instrument called SKA1-Mid. All 3 telescopes will operate independently as part of a single SKA Observatory. In Phase 2 the dish array (Figure 1) in South Africa will be built out further, totalling about 3000 dishes and including stations in the partner African countries, delivering baselines as long as 3000 km. The low frequency aperture array will be expanded in Australia to the full SKA collecting area.  In Phase 2 the dense, or mid-frequency, aperture array (Figure 1) will be built in South Africa, pending successful review of this technology as part of the SKA Advanced Instrumentation Programme (AIP). 

%Example how to use figure environment
\begin{figure}[htb]
\centering
\includegraphics[width=0.7\textwidth]{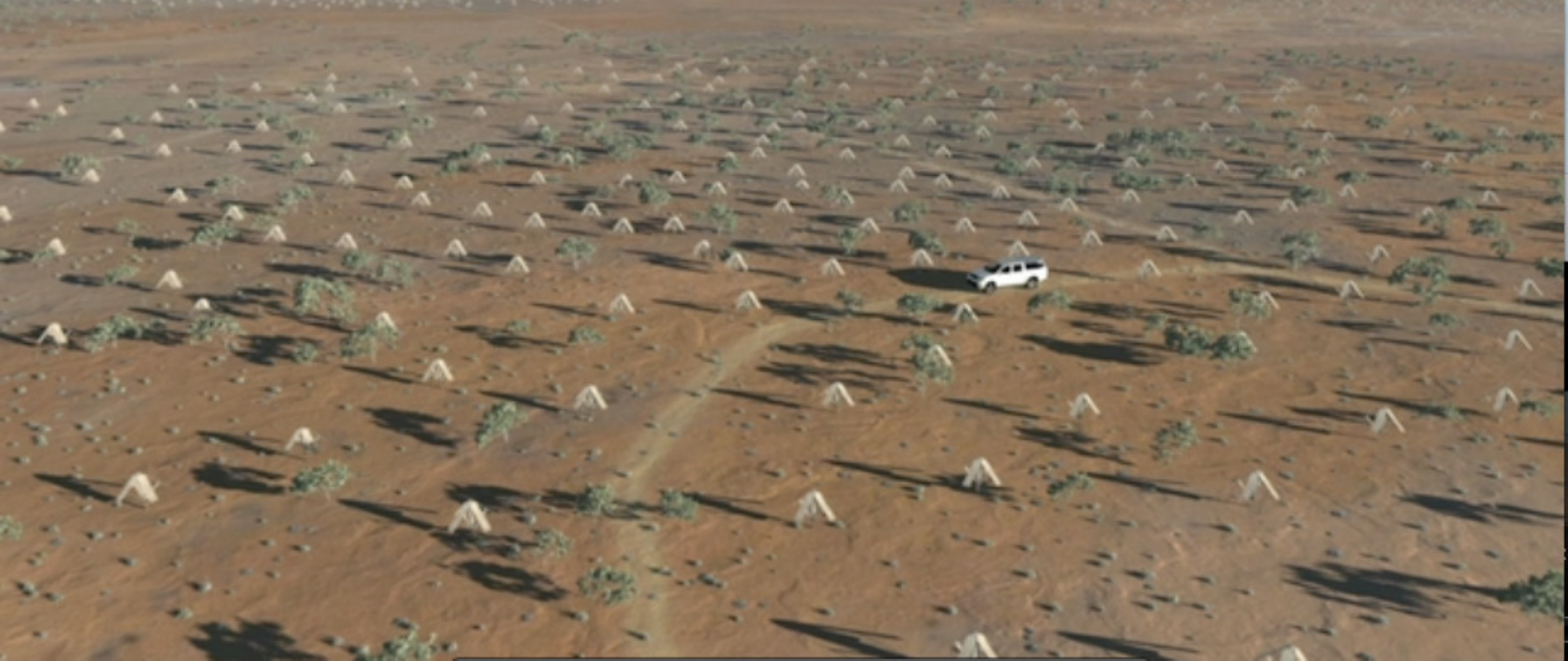} 
\includegraphics[width=0.7\textwidth]{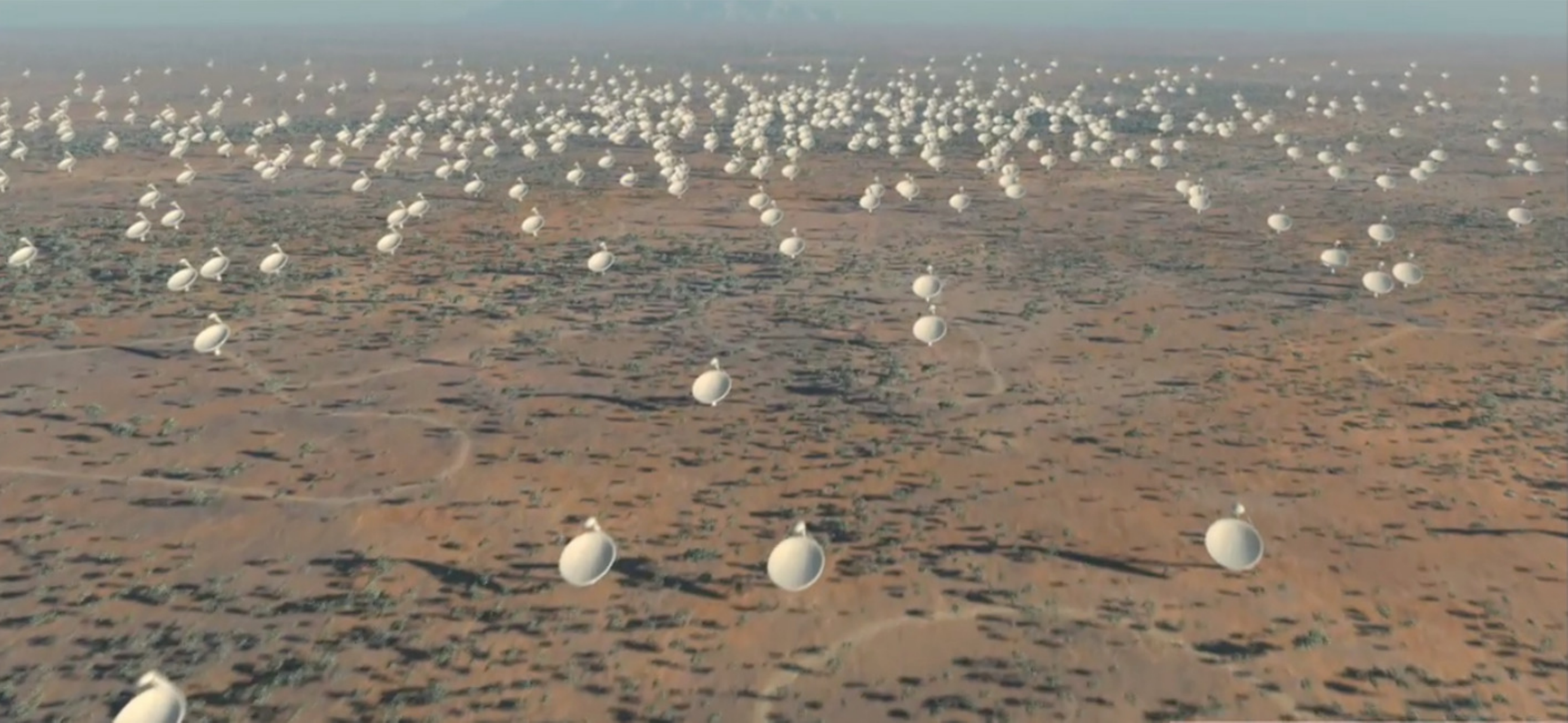} 
\includegraphics[width=0.7\textwidth]{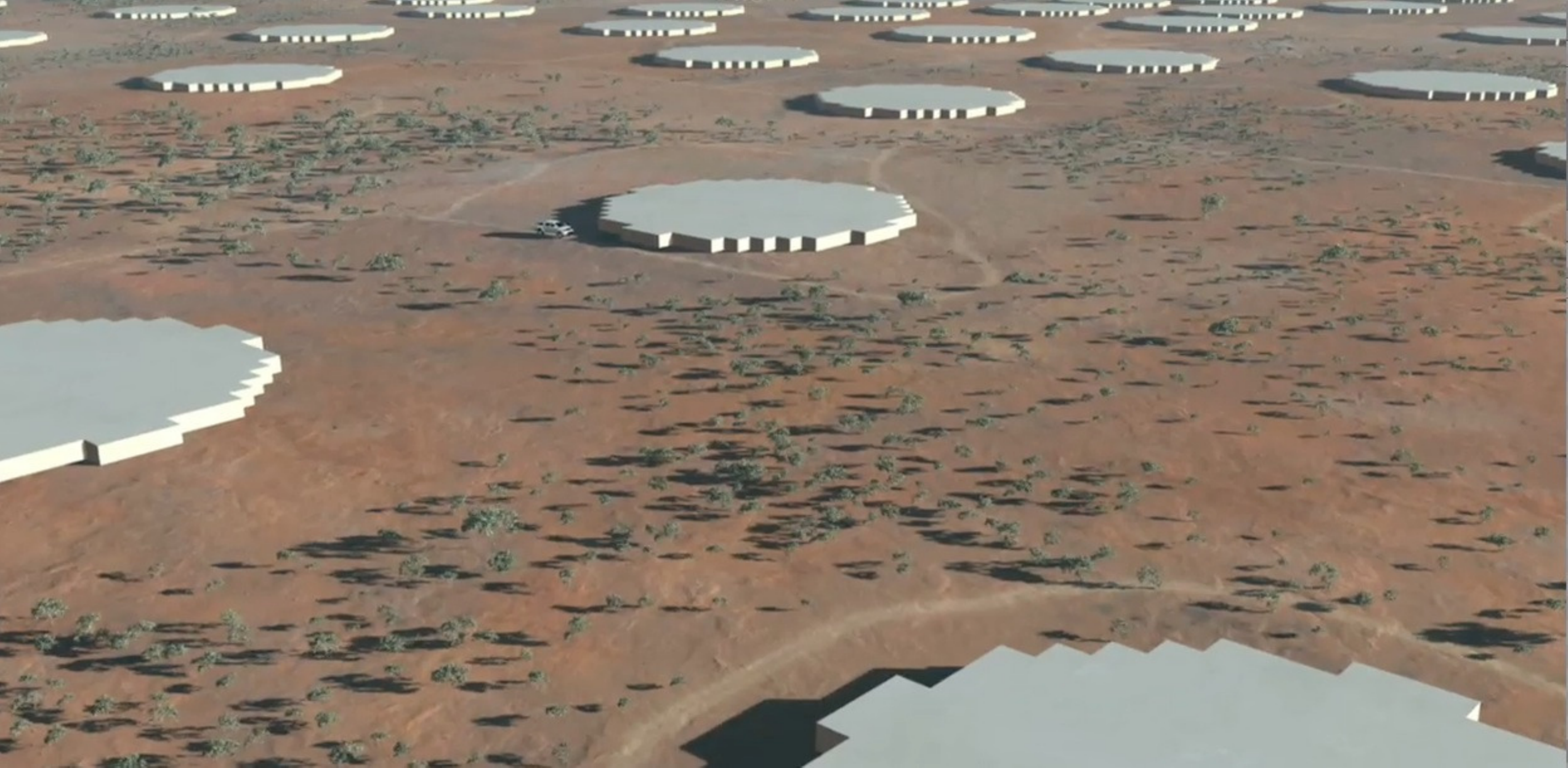}
\caption{Artist conception of the SKA low-frequency aperture array (top), dish array (middle), and dense mid-frequency aperture array (bottom).}\label{figure1}
\end{figure}

\subsection{Preconstruction and the Baseline Design}

Over the next four years, 2013 -- 2016, the SKA Organisation is in the preconstruction and detailed-design phase. The goal of this phase is to prepare the SKA project at element level to the point where construction of SKA Phase 1 can begin. In March of this year the SKA Organisation released a Request for Proposals for preconstruction design work. The telescope is broken down into Elements for this work, largely following a traditional radio telescope work breakdown structure. These Elements are: \\ 
i) Dish (including phased array feeds), \\
ii) Low Frequency Aperture Array, \\
iii) Telescope Manager, \\
iv) Central Signal Processor (including correlator and pulsar timing), \\
v) Science Data Processor, \\
vi) Signal And Data Transport, \\
vii) Assembly, Integration and Verification, \\
viii) Infrastructure, and \\
 ix) Advanced Instrumentation Programme (dense aperture arrays and wide band single pixel feeds). \\
The design work will be fully funded by the bidding consortia which responds to the Request for Proposals. The winning work package consortia are expected to be announced by October 2013, in time for the kickoff Engineering Meeting in Manchester.  
Stage 1 of the preconstruction work is to get the element to Preliminary Design Review (PDR) by October 2014. Stage 2 is to go from PDR  to construction readiness (or Critical Design Review) by the second half of 2016. Tender and procurement is expected in 2017, and SKA Phase 1 construction to begin in 2018.

As part of the Request for Proposals the SKA Organisation has developed a baseline design for the telescope.The baseline design is a description of the basic 3-telescope, 2-system dual-site architecture (SKA1-Low, SKA1-Survey, SKA1-Mid). The baseline design is summarised in Table 1 (see Dewdney et al.~2013 for full description).
The aim was to produce a design that emphasises the capability to do key Phase 1 science while preserving flexibility for all types of observations. 
This baseline design is considered the starting point for the preliminary design phase and is not the final design. Controlled changes are permissible and will result from responses to the Request for Proposals, outcomes from Science Assessment Workshops run by the SKA Organisation, and design work by the consortia during preconstruction. Major design changes will have to be extremely well motivated and the final design will be based on cost analysis, carried out as well considered trade-offs. 

\begin{table}
\begin{tabular}{llrrrr} \hline 
 & & SKA1-Low & SKA1-Survey & SKA1-Mid \\ \hline
Frequency Range & MHz & 50 -- 350 & 650 --1670 (initially) &  350 -- 1670 (initially)\\
 Fiducial Frequency & MHz & 110	& 1400 & 1400	\\
 Aeff/Tsys & m$^2$ /  K  & 1000 & 390 & 1600	\\
 FoV & deg$^2$ & 27 & 18 & 0.68\\
 Survey Speed FoM & deg$^2$ m$^4$ / K$^2$ & 2.70 $\times 10^7$ &  2.75 $\times 10^6$ & 1.74 $\times 10^6$		\\
 Maximum Baseline & km & 60 & 50	& 200 \\
 Resolution & arcsec & 7 & 0.9 & 0.2 \\
Instantaneous Bandwidth & MHz & 300 & 500	& 770 \\
Frequency Resolution & kHz & 1 & 	1.95 & 	3.9 \\ \hline
 
\end{tabular}
\caption{Summary of the baseline design of SKA Phase 1. The Aeff/Tsys, survey speed, field of view and resolution numbers given are for the fiducial frequencies as shown. }
\end{table}

\subsection{SKA Science Working Group and Science Teams}

A new SKA Science Working Group structure was implemented in the beginning of 2013. The SKA Science Working Group is led by the SKA Science Director. Working under the Science Director are the SKA Project Scientist(s) and Deputy Project Scientist. Also in the Science Working Group are regional project scientists and a delegate or chair from each of the Science Teams. There are currently eight Science Teams, covering: i) Pulsars for Fundamental Physics, ii) Epoch of Reionization and the Dark Ages, iii) H~I Surveys for Galaxy Evolution, iv) Continuum Surveys, v) Cosmology, vi) Cradle of Life / Astrobiology, vii) Cosmic Magnetic Fields, and viii) Transients. The Science Working Group (SWG) consists of the key science personnel and coordinate the work of the Science Teams. The Science Teams represent the broader community and thus can have a much larger membership. They are expected to: 1)  provide guidance to the SKA Organisation / SWG on the science drivers from each respective area, 2) Provide assistance in developing technical requirements from the science goals and requirements, and 3) Serve as SKA liaisons to the broader community. Science Team membership, a responsibility of the Science Team chair(s), is open to any researcher with a science interest in the SKA and a willingness to contribute an appropriate level of effort toward SKA science needs. 

The work of the Science Teams this year has included evaluating and providing feedback on the baseline design to the SKA Organisation. As part of this, Science Assessment Workshops have been held at the SKA Organisation headquarters at Jodrell Bank Observatory. A Science Assessment Workshop  on the Epoch of Reionization was held in March 2013, and workshops on pulsar, continuum and more ``local" H~I science are planned\footnote{Summaries of the Science Assessment Workshops will be available online at http://www.skatelescope.org}. Looking further ahead, the Science Teams are expected to assist in revising and updating the SKA Science Case.  As the SKA moves to Preliminary Design Review over the next year, and then Critical Design Review in 2016, support from the Science Teams is crucial to ensure that the SKA Phase 1 is an instrument capable of delivering transformational science. 

\vspace{5mm}
{\noindent \footnotesize Part of this research was carried out at the Jet Propulsion Laboratory, California Institute of Technology, under a contract with the National Aeronautics and Space Administration.}

\end{document}